# Guest-induced structural deformation in Cu-based Metal-Organic Framework upon hydrocarbon adsorption

Azahara Luna-Triguero,*[†][a,b] Eduardo Andres-Garcia,[†][c,d] Pedro Leo,*[c,e] Willy Rook,[c] Freek Kapteijn[c]

[a] Energy Technology, Department of Mechanical Engineering, Eindhoven University of Technology, P.O. Box 513, 5600 MB Eindhoven, The Netherlands.
[b] Eindhoven Institute for Renewable Energy Systems (EIRES), Eindhoven University of Technology, PO Box 513, Eindhoven 5600 MB, The Netherlands
[c] Catalysis Engineering, ChemE, TUDelft, Van der Maasweg 9, 2629 HZ Delft, The Netherlands.
[d] Instituto de Ciencia Molecular (ICMol), Universidad de Valencia, c/Catedrático José Beltrán, 2, Paterna, 46980, Spain.
[e] Department of Chemical and Environmental Technology, Rey Juan Carlos University, Calle Tulipán s/n, 28933 Móstoles, Spain.

E-mail: a.luna.triguero@tue.nl, pedro.leo@urjc.es

## Abstract

*In a world where capture and separation processes represent above 10% of global energy consumption, novel porous materials, such as Metal-Organic Frameworks (MOFs) used in adsorption-based processes are a promising alternative to dethrone the high-energy-demanding distillation. Shape and size tailor-made pores in combination with Lewis acidic sites can enhance the adsorbate-adsorbent interactions. Understanding the underlying mechanisms of adsorption is essential to designing and optimizing capture and separation processes. Herein, we analyze the adsorption behaviour of light hydrocarbons (methane, ethane, ethylene, propane, and propylene) in two synthesized copper-based MOFs, Cu-MOF-74 and URJC-1. The experimental and computational adsorption curves reveal a limited effect of the exposed metal centers on the olefins. The lower interaction Cu-olefin is also reflected in the calculated enthalpy of adsorption and binding geometries. Moreover, the diamond-shaped pores' deformation upon external stimuli is first reported in URJC-1. This phenomenon is highlighted as the key to understanding the adsorbent's responsive mechanisms and potential in future industrial applications.*

## Keywords

## Introduction

Natural Gas (NG) demand accounts for the highest fuel growth rate, and its consumption is expected to surpass coal's in 2030.[1] Although NG is generally considered clean energy, it is not free of impurities – such as water, $CO_2$ and other hydrocarbons – that need to be removed by energy intensive separation processes;[2] consequently, a large amount of the world's current natural gas reserves are not available for being economically unprofitable.[3] In addition, the contribution of methane to the greenhouse effect should not be underestimated due to its elevated global warming potential (GWP).[4] Propane/propylene separation is considered the most challenging separation in the chemical engineering industry for two main reasons: i) the similarities of the components in the mixture; ii) the high value of the feedstock involved in the process.[5-7] Both ethylene and propylene are two of the most important feedstocks in the chemical industry, with multiple applications in the refinery, and used as building block in the production of some of the most common chemicals and polymers. Although these alkenes are usually obtained in globally equimolar mixtures with their corresponding alkane (ethane and propane), industrial purity requirements demand new energy efficient separation techniques. Adsorption-based separation using porous materials is the technology to reach the required high purity (99.95 mol% ethylene;[8] 99.5 mol% propylene),[9] avoiding the high capital/operational costs of the energy-demanding cryogenic distillation.[10]

Metal-Organic frameworks (MOFs), a sub-class of porous coordination polymers (PCPs), are hybrid organic-inorganic porous materials well known for their promising structural properties such as high surface area, pore volume, and relatively high thermal and chemical stability. Accordingly, MOFs are being currently explored in a wide range of applications such as gas sequestration and separation,[11-12] energy storage,[13] drug or biomolecule release,[14-15] and heterogeneous catalysis.[16]

Among the above mentioned properties, it is worth highlighting the chemical and functional tunability; the abundant number of organic ligands, metals and metallic clusters, crystallization conditions, and post-synthesis modifications allow almost infinite combinations.[17-21]

Currently, the Cambridge structural database is the platform with more indexed MOFs, and the number of synthesized materials is growing exponentially.[22-23] The vast majority of reported MOFs can be classified as the second generation; robust frameworks with permanent porosity considered as rigid porous materials. On the other hand, structural flexibility is a desirable property that can be exploited for different applications and can theoretically be used for gas separation,[24-28] although this has been rarely explored for mixtures since their performance is hard to measure and predict.[29-30] MOFs presenting framework flexibility, reversible structural changes upon external stimuli, are known as soft porous crystals (SPCs), the third generation of MOFs.[31] Only about 100 structures are reported to exhibit reversible phase transitions; however, it is a highly influential factor in understanding the underlying adsorption and diffusion mechanisms and performance prediction.[32-33]

Here we compare the adsorption mechanisms of two Cu-based MOFs with one-dimensional diffusivity channels, Cu-MOF-74 and URJC-1. The adsorption of light hydrocarbons, methane, ethane, propane, ethene, and propene in Cu-MOF-74 and URJC-1, based on experimental data and computational results is investigated. Synthesis of the materials, structural characterization, and adsorption isotherms are performed and analyzed by calculating the enthalpy of adsorption, binding geometries, and structural optimizations.

Cu-MOF-74 is a variant of the M-MOF-74 family; this family is formed by a set of well-known structures with open M(II) sites with M=Co, Fe, Mg, Mn, Ni, Zn, and Cu.[34-39] They consist of large hexagonal pores of 10-12 Å diameter where the metal clusters propagate in the *c*-axis. This family has been reported for gas separations such as olefin/paraffin,[40-42] carbon dioxide/methane,[43-44] and acetylene purification.[45-47] The performance of M-MOF-74 for separation is linked to the strong interaction of its exposed metal centers with certain molecules through π-complexation. While other members of this family have been extensively studied, only a few studies include Cu-MOF-74.[43, 48-49]

URJC-1 is a recently reported MOF proposed for different catalytic applications.[50-51] This MOF is characterized by its diamond-shaped pores of about 5 Å and accessible Lewis sites. URJC-1 has open Cu(II) sites pentacoordinated with the nitrogen atoms of the tetrazole and imidazole rings.

Ultimately, an unexpected behaviour of URJC-1 is reported. This highly stable MOF presents an induced framework deformation during the adsorption process. This phenomenon is evidenced by the structure optimization under different conditions and supported by the obtained adsorption isotherms of the hydrocarbons and comparison with the experimental results. An accurate description of the URJC-1 phases is reported for the first time and the obtained structures are well characterized.

## Methodology

### Synthesis of Cu-MOF-74 and URJC-1

All analytical reagents were commercial products and they were used without further purification.

The synthesis procedure of Cu-MOF-74 was slightly modified from the literature.[39] A mixture of 2,5-dihydroxyterephthalic acid (11.2 mmol) and $Cu(NO_3)_2·3H_2O$ (24.6 mmol) were added over a 20:1 (v/v) solution of DMF and 2-propanol (250 mL). The reaction vial was capped tightly and placed in an oven at 80 °C during 18 h.

URJC-1 was synthesized following the procedure previously reported.[50] In a typical synthesis, the material was prepared by mixing 1H-imidazol-4,5-tetrazole as the organic ligand and $Cu(NO_3)_2·3H_2O$ as the inorganic source in an acidified solution of N,N´-dimethylformamide and acetonitrile as solvents (6 mL, 3:3, v/v). This mixture was heated at 150 °C for 20 h using a heating ratio of 1.5 °C/min. The representation of the synthesized structures can be found in Figure 1.

### Chemical characterization

X-ray powder diffraction (XRD) patterns, Figures S1 and S2 in the Supporting Information (SI), were acquired on a PHILIPS X'PERT diffractometer using Cu *Kα* radiation. The data were recorded from 5 to 50° (2ϴ) with a resolution of 0.01°. Fourier transform-infrared spectra (FT-IR) on powdered samples were carried out on a Varian 3100 Excalibur Series spectrometer with a resolution of 4cm$^{-1}$ and 64 scans coupled to an MKII Golden Gate Single Reflection ATR system to acquire spectra in Attenuated Total Reflectance mode.

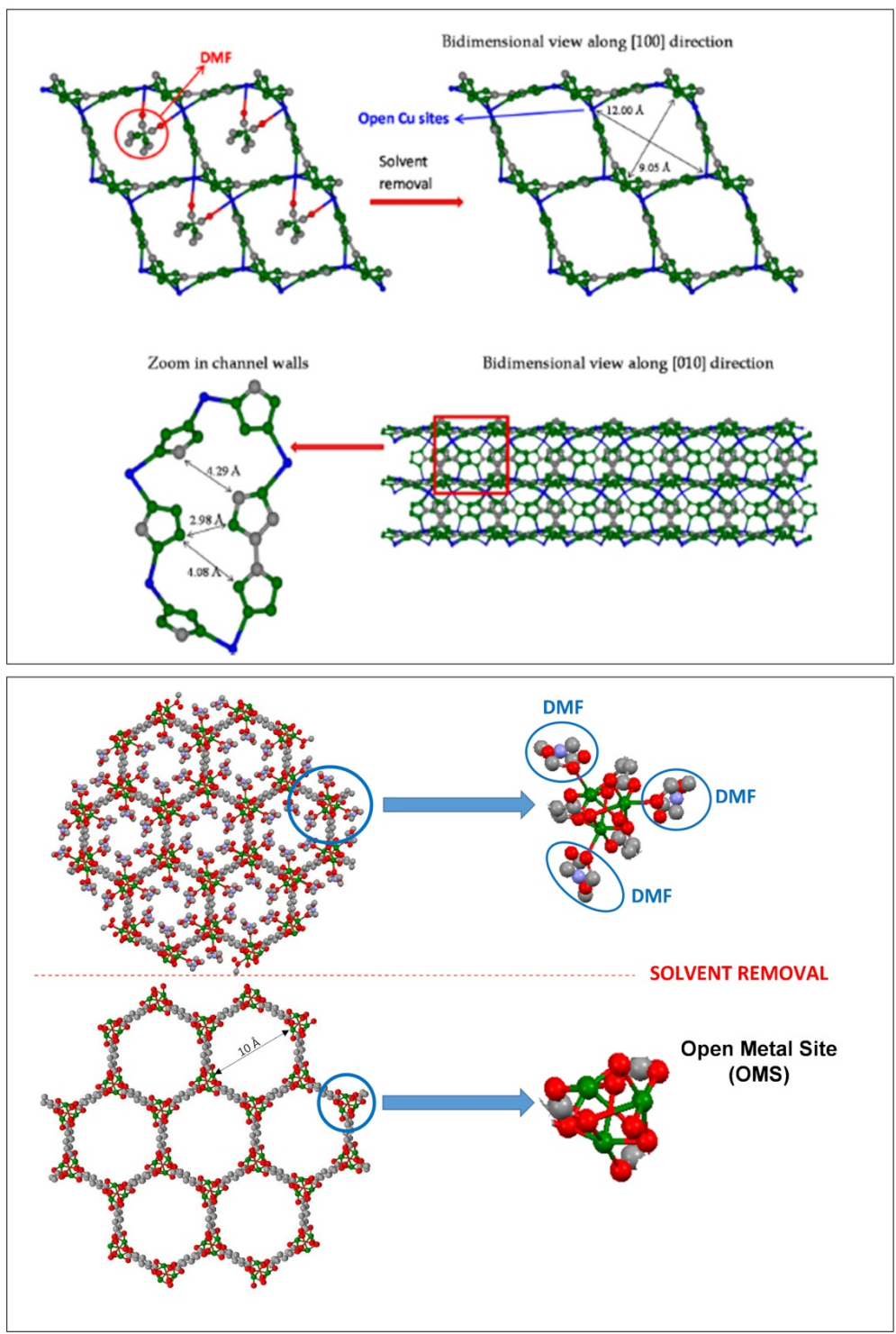

**Figure 1**. Schematic representation of the synthesized structures URJC-1 (*top*) and Cu-MOF-74 (*bottom*).

**Single-gas adsorption measurements**

Both adsorbents were analysed by low-pressure physisorption. Adsorption and desorption hydrocarbons isotherms were collected in a Tristar II 3020 (Micromeritics). Experiments were conducted at 25 °C (298 K) for methane, ethane and ethylene and at 30 °C (303 K) for propane and propylene. Nitrogen adsorption-desorption isotherms at

77 K were measured using AutoSorb equipment (Quantachrome Instruments). Both MOF samples were degassed under vacuum at 150 °C for 16 h. Nitrogen adsorption isotherms can be found in Figure S3 in the SI.

**Computational details**

Equilibrium adsorption isotherms were calculated using Monte Carlo simulations in the grand canonical ensemble (GCMC) with the aim of comparing them with the experimental results. Simulations were performed using RASPA software.[52-53] Each point of the adsorption isotherms is obtained after $10^5$ MC cycles, and the production runs were performed after $10^4$ MC equilibration cycles. The frameworks are considered rigid structures during the adsorption, with the atoms placed at the crystallographic positions. URJC-1 was optimized using classical structural minimizations with an empty and half-loaded structure with different adsorbates. The geometry optimizations were carried out using NPT ensemble allowing independent variation of the cell lengths and the angles. The configurations were selected from the optimization of the original crystal structure, and the minimizations were performed with loaded molecules corresponding to half of the saturation conditions according to the experimental data following the reported methodology.[54-55] We performed structural characterization of the original crystal structure and the obtained minimizations calculating the pore size distribution (PSD), surface area, and pore volume. Moreover, we carried out energy minimizations using Baker's method [56] in canonical ensemble (NVT) to obtain the binding geometries calculated with a single molecule of the adsorbates in Cu-MOF-74 and URJC-1.

The non-bonded interactions consist of adsorbate-adsorbent and adsorbate-adsorbate Van der Waals and electrostatic interactions. The Van der Waals interactions are modeled using 12-6 Lennard-Jones potential. The Lennard-Jones parameters for the framework atoms are taken from DREIDING and UFF for the metal atoms.[57-58] The models of the adsorbates are taken from literature. Paraffins are described using a non-charged pseudoatom model where each $CH_n$ group is considered as a single interaction center.[59-61] For olefins, a point charge model is used where partial charges are located in the $CH_n\_sp^2$ groups and a point charge is located between the carbon atoms linked by the double bond.[62-63] Lorentz-Berthelot mixing rules are applied to account for the cross interactions. A set of effective point charges is used for the framework atoms. The charges were obtained using EQeq method based on Ewal sums.[64] The resulting charges are listed on Table S1 and the atoms labels shown in Figure S4 of the SI.

**Results and discussion**

To understand the adsorption mechanisms at the microscopic level, we measured volumetric adsorption isotherms of light hydrocarbons in URJC-1 and Cu-MOF-74.

Figures 2 and 3 illustrate the behaviour of the two copper adsorbents URJC-1 and Cu-MOF-74 in contact with single-gas hydrocarbon atmospheres under isothermal conditions. Although URJC-1 presents similar adsorption profiles for both alkane/alkene doublets: Langmuir type isotherms, adsorption capacities in a range 2.94-2.96 mmol g$^{-1}$ (C2) and 2.38-2.58 mmol g$^{-1}$ (C3) and lack of hysteresis loops; remarkably, C2s capacity is slightly larger than C3s, as consequence of adsorption-induced framework deformation and the associated changes in the MOF properties later discussed. However, the analysis of adsorption kinetics reveals the diffusional impediments for the alkanes, potentially resulting in a preferential adsorption of olefins in an equimolar mixture.

MOF-74(Cu) exhibits different profiles for the C2/C2= and C3/C3= couples: despite propane and propylene isotherms resemble the URJC-1 results, the adsorption capacities are higher and over a broader pressure range, showing a clearer selectivity towards the alkene. Ethane/ethylene curves almost coincide, with a less steep slope adsorption branch than in URJC-1, hindering an alkane/alkene separation, but allowing a good C2-C3 separation, in that pressure range.

Figure 3 compares C1, C2, and C3 alkanes adsorption for both MOFs. Methane adsorption capacity is lower than for the other hydrocarbons, particularly in MOF-74(Cu), where the *type III* isotherm is a consequence of low adsorbate-adsorbent interactions. These differences in the adsorption branch, especially in the low-pressure region, evidence a promising potential in adsorptive gas separation processes.

Figure 4 compares the experimental results for methane, ethane and propane adsorption-desorption uptake and the computed adsorption isotherms obtained using the crystallographic (as-synthesized) structures. A good agreement between the experimental and simulation adsorption data for Cu-MOF-74 is obtained, with a really accurate result for methane and ethane at 298 K and a slight overestimation for propane adsorption at saturation conditions and 303 K.

As in the case of the paraffins, the computed adsorption isotherms of ethene and propene in Cu-MOF-74 are in good agreement with the experimental data (Figure 5). Cu-MOF-74 has a remarkably different behaviour compared to that observed for other M-MOF-74 structures. It is well studied and reported that M-MOF-74 is a good candidate for capturing saturated and unsaturated hydrocarbons; the big channels of the MOFs allow high capacity at ambient conditions.[65-66] Because of the π-complexation between the open metal sites (OMS) and the double bond of the olefins, this family of MOFs has also been proposed for alkane/alkene separation.[42, 67-70] In this regard, Cu-MOF-74 is an exception; the interaction between the Cu atom and the olefins is weaker than for the other metals, *i.e.,* Co, Fe, Ni, Mg, Mn, and Zn. This behaviour is reflected in the adsorption isotherms of ethene and propene, where the onset pressure in the Cu-based MOF is one order of magnitude higher than for the other metals but shows similar trends in the adsorption of paraffins (Figure S5). Another difference between Cu-MOF-74 and the remaining M(II)-MOF-74 family members is the capacity. Cu-MOF-74 shows lower loading values at similar conditions, which is consistent with its slightly smaller pore volume. This phenomenon is especially pronounced in the case of olefins (Figure S5).

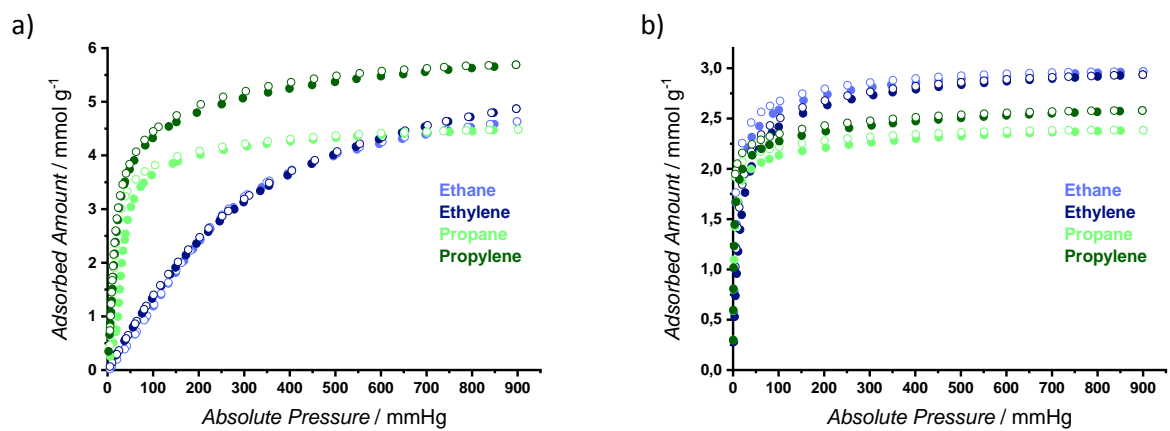

**Figure 2.** Low-pressure volumetric adsorption/desorption isotherms of ethane (light blue) and ethylene (dark blue) at 298 K and propane (light green) and propylene (dark green) at 303 K. Adsorption (solid symbols) and desorption (open symbols) in Cu-MOF-74 (a) and URJC-1 (b).

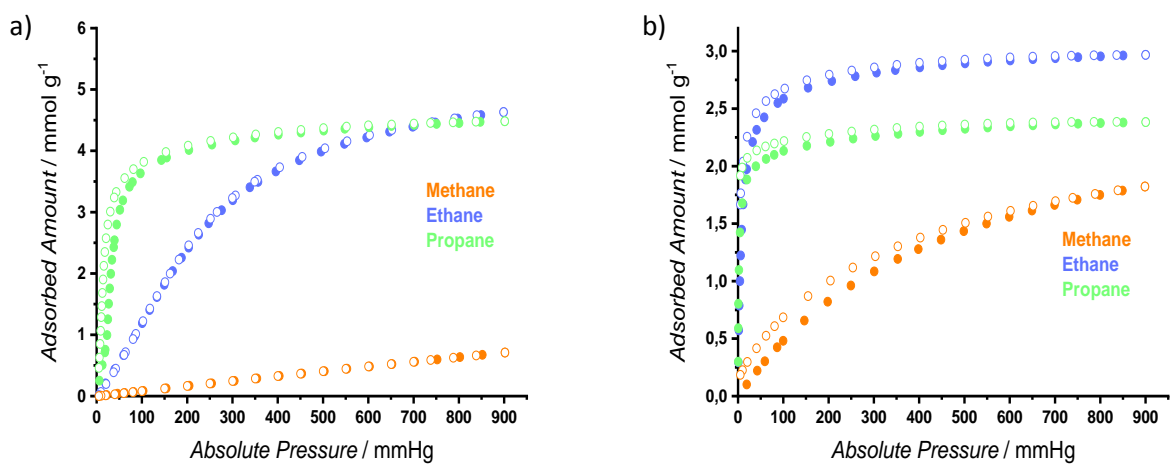

**Figure 3.** Low-pressure volumetric adsorption/desorption isotherms of methane (orange), ethane (blue), and propane (green), at 298 K, 298 K, and 303 K respectively. Adsorption (solid symbols) and desorption (open symbols) in Cu-MOF-74 (a) and URJC-1 (b).

On the other hand, the calculated adsorption of methane, ethane, and propane in URJC-1 show a systematic overestimation, especially pronounced for propane. To understand the behaviour of this MOF, structural optimizations of URJC-1 are conducted, empty and pre-loaded with adsorbates up to half of the loading. The calculations are made at intermediate loadings to ensure that the resulting optimized structures are not an artifact of the overestimated saturation capacity. From the energy minimizations two structures are obtained, the empty (URJC-1-e) and loaded (URJC-1-A) structures. The adsorption isotherms of methane, ethane, and propane are computed with both optimized structures. As shown in Figure 4, the adsorption curves obtained using URJC-1-e are shifted in the pressure range, overestimating the loading over the entire range (open symbols in Figure 4). Meanwhile, the computed adsorption isotherms of methane and ethane in URJC-1-A are in good agreement with the experimental measurements (solid symbols in Figure 4).

Given the structural deformation occurred during optimization in presence of the adsorbates and the deviation of the adsorption curves derivated from them, the same approach is used for alkenes. We performed structural optimization of the pre-loaded URJC-1 with alkenes at half of the experimental saturation. As a result, a third structure is obtained, URJC-1-B. Adsorption isotherms of ethene and propene were computed for each structure and compared with the experimental measurements (Figure 5). The adsorption isotherms show an overestimation of the entire curve from 1 to 1.5 mol/kg for ethene and up to 2 mol/kg for propene in URJC-1-e and URJC-1-A. A good agreement, however, was obtained between the experimental observations and calculated adsorption isotherms in URJC-1-B (Figure 5, full symbols).

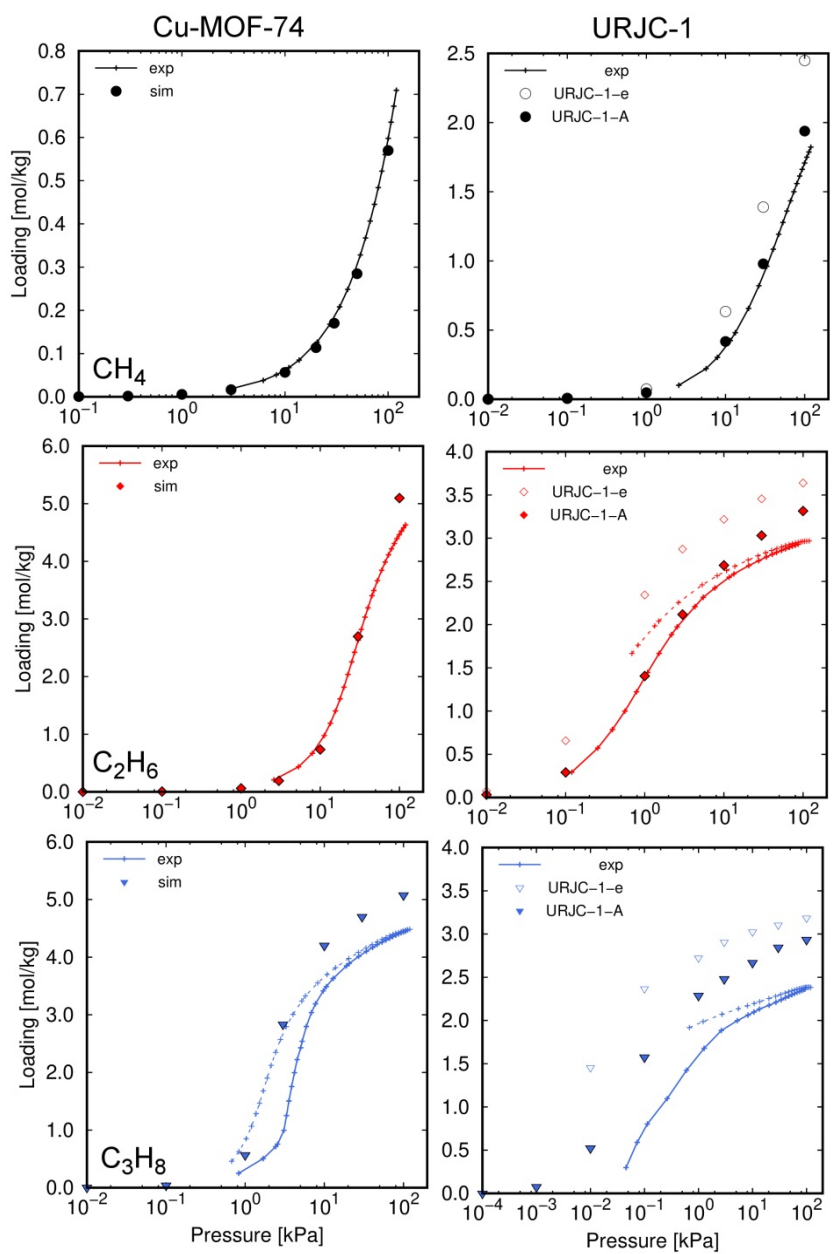

**Figure 4.** Single component adsorption isotherms of methane (black circles), ethane (red diamonds) at 298 K, and propane (blue triangles) at 303 K in Cu-MOF-74 and URJC-1. Comparison of experimental adsorption (solid lines) and desorption branches (dashed lines), and calculated adsorption (symbols)

In addition to the adsorption isotherms, the adsorption enthalpy as a function of the loading using the fluctuation method is calculated.[71] The enthalpy of adsorption follows the expected hierarchy, showing larger interaction with the longer molecules, and the lowest with methane (Figure 6). The absolute values for the same adsorbate are higher for URJC-1 due to the difference in pore size. The higher interaction between the exposed metal site and the olefins is well reported in the M-MOF-74 analogs and attributed to the π-complexation.[72-73] This phenomenon is also apparent in Cu-based open metal site MOFs.[74-77] However, neither of the MOFs shows evidence of the effect of the open metal sites over the olefins.

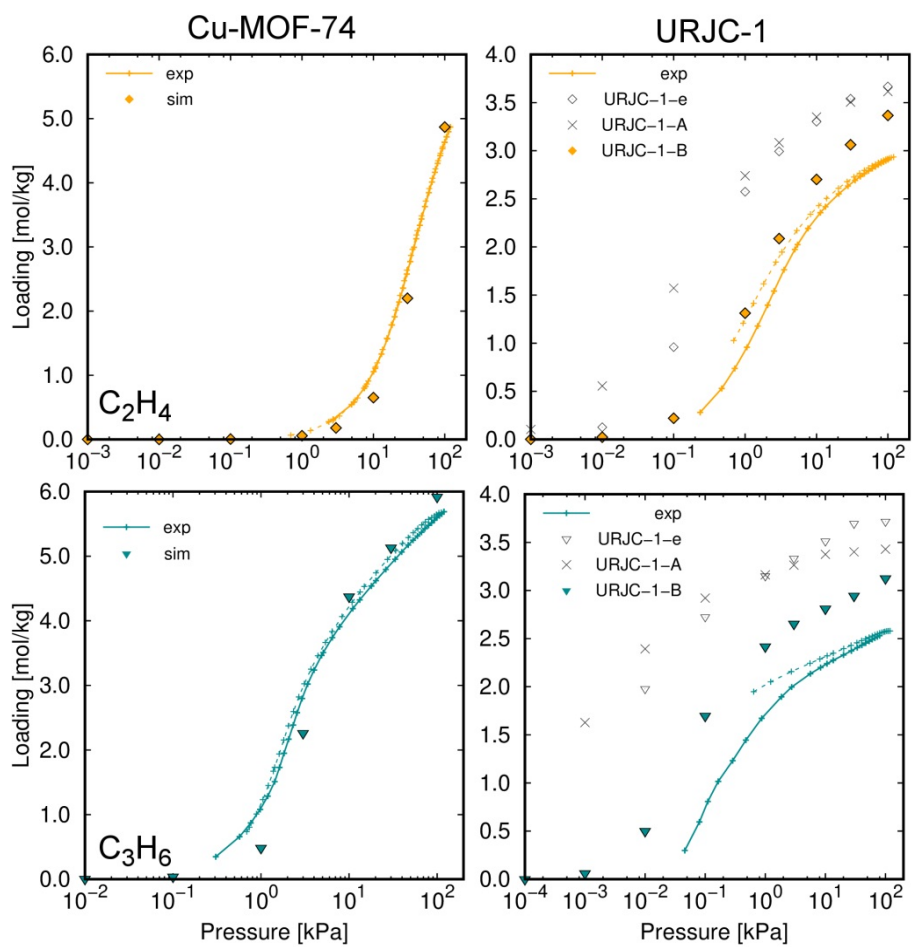

**Figure 5.** Single component adsorption isotherms of ethene (yellow diamonds) at 298 K and propene (turquoise triangles) at 303 K in Cu-MOF-74 and URJC-1. Comparison of experimental adsorption (solid lines), desorption branches (dashed lines), and calculated adsorption (symbols).

The binding geometries also show the weaker Cu atoms-double bond interaction observed in the structures under study (Figure 7). Three parameters are defined to quantify and analyze the binding geometries; the distance between the closest carbon atom to the metal center, $l_1$; the second closest C-Cu distance, $l_2$, and the angle between the C-C bond of the adsorbate and the open metal site, α. The parameters are schematically represented in Figure 7. The hydrocarbon-Cu-MOF-74 binding geometries show considerable longer distances than the other MOFs of the series. This phenomenon was also observed for hydrogen adsorption[78] and carbon dioxide[43], where the Cu-adsorbate interaction is weaker than for the M-MOF-74 analogs. The distance Cu-methane, 3.94 Å, is similar to the reported for Mg-methane, 4 Å.[79]

In URJC-1, the molecules are adsorbed in the center of the diffusion channel giving larger Cu-C distances. This phenomenon is caused by the narrow pores that make the center of the cavity the only available adsorption site (Figure S6). Despite the significant difference in the distances, the orientations of the molecules are similar. It is easy to identify two different behaviour groups; the double bond of ethene and propene faces the Cu atom, and α is between 82-85 degrees showing angles typically attributed to the π-complexation. For paraffins, one of the extreme carbon atoms is near the Cu-center forming Cu-C-C angles of 106-110 degrees. These angles are in agreement with those previously reported in the literature.[80] Olefins are placed with the double bond parallel to the metal cluster, while paraffins bind with one extreme carbon atom pointing toward the metal.[45]

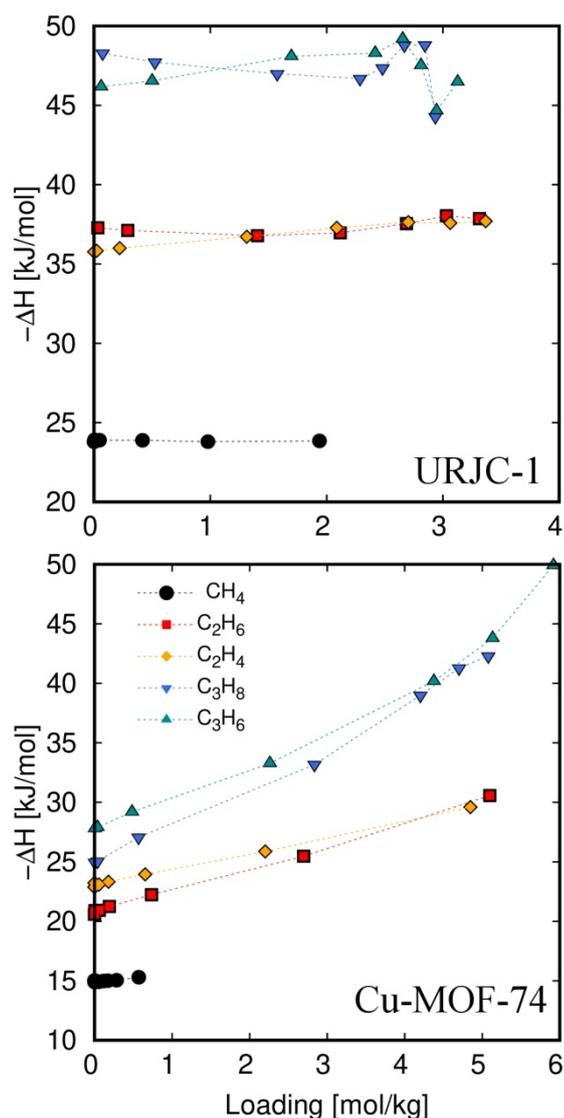

**Figure 6.** Enthalpy of adsorption as a function of the loading of methane, ethane, ethene, propane, and propene in URJC-1 and Cu-MOF-74.

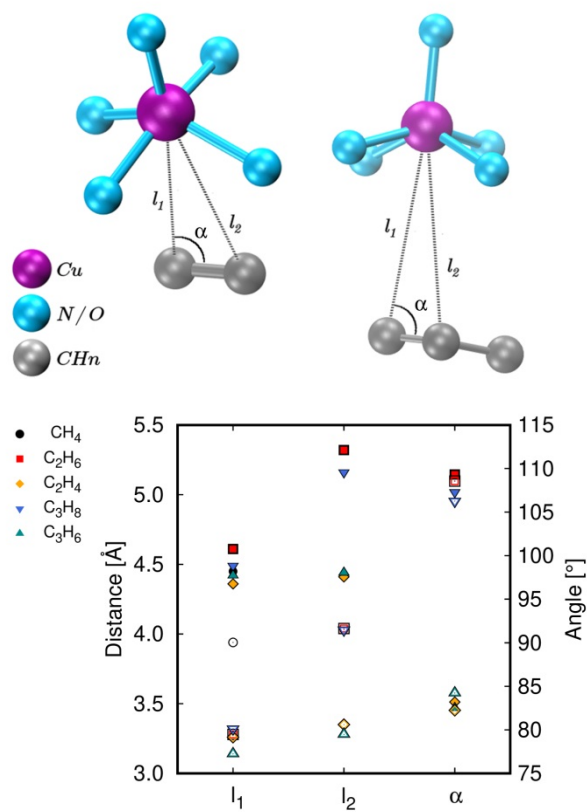

**Figure 7.** Schematic representation (*top*) and binding geometry parameters (*bottom*) of $CH_n$-Cu atoms in URJC-1 (*solid symbols*) and Cu-MOF-74 (*open symbols*).

From this point, the focus is on the structural analysis and characterization of the optimized URJC-1 to understand the consequences of the framework deformation triggered by the adsorbates. The results concerning URJC-1 are calculated with the structure that reproduces the adsorption isotherms, i.e., URJC-1-A for methane, ethane, and propane, and URCJ-1-B for ethene and propene.

The optimized empty structure (URJC-1-e) is practically identical to the crystallographic structure, maintaining pore size, volume, and symmetry, becoming the reference structure. The structures show an almost negligible variation of the lattice parameters, cell lengths, and angles (Table S2), which cannot explain the difference observed in the adsorption. Therefore, the change in the shape of the pores becomes the focus of attention. A set of distances are defined and compared to account for these differences (Figure 8). The triangle formed by the in-plane copper atoms defines $d_1$, $d_2$, and $d_3$ distances. URJC-1-A shows a similar increase in the three distances compared to the empty structure. A similar increase is appreciated in $d_3$ URJC-1-B, while the $d_1$ increment is twice the obtained for URJC-1-A, and $d_2$ shows a slight contraction. To understand the implications of the mentioned incremental distances, $d_4$ and $d_5$ are also analyzed. $d_4$ is defined as the distance between the face-to-face Cu atoms, and $d_5$ is the distance between the faced nitrogen atoms of the tetrazole ring. For URJC-1-A, $d_4$ is shortened but close to the reference distance, while $d_5$ is elongated in the same proportions as $d_1$-$d_3$. The most significant difference is shown in the elongation of $d_4$ by URJC-1-B, which is compensated by a contraction in $d_5$. Finally, the pore deformation can be observed in the torsion angle N-C-C-N, which defines the angle between the planes of the tetrazole and imidazole rings (Figure S7). The distortion of the diamond-shaped pore is evidenced by minor variations in the configuration of the pores that lead to considerable differences in the adsorption behaviour (Figures 4 and 5).

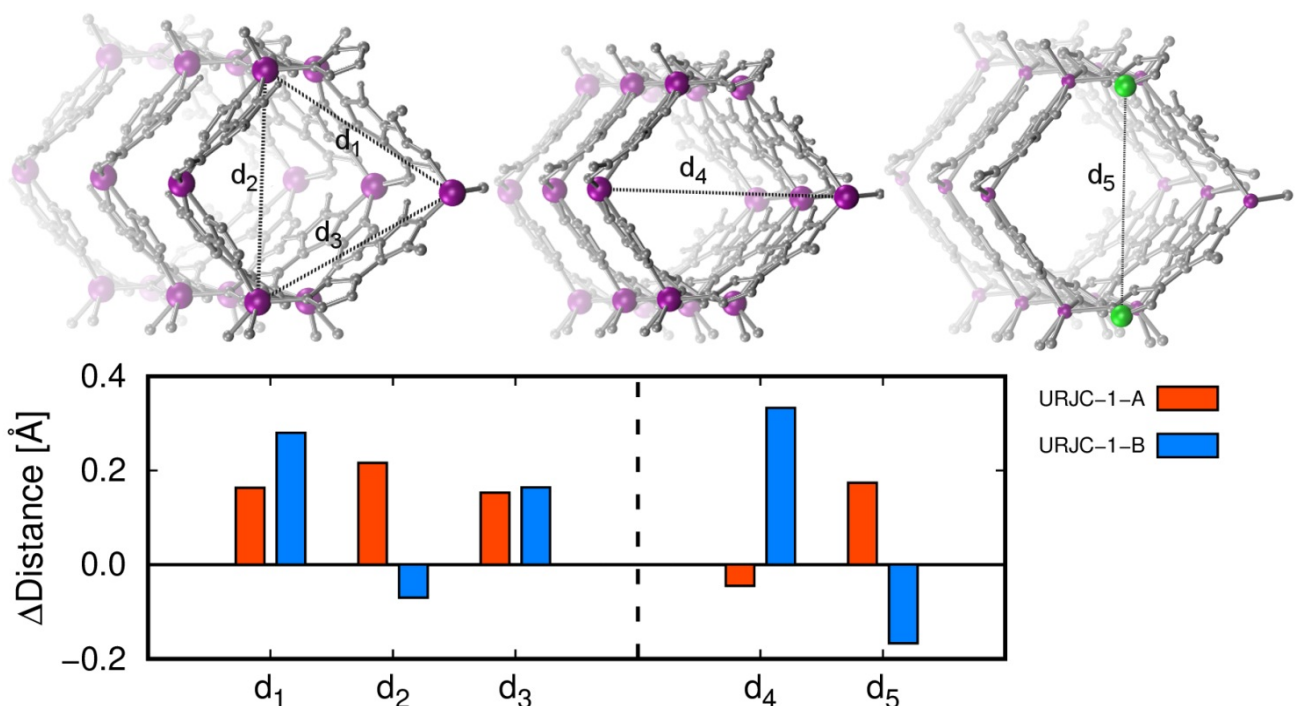

**Figure 8.** Schematic representation of atomic connectivity and distance definitions. Copper atoms in violet, nitrogen atoms in green, and the remaining atoms in grey. Incremental distance between the reference structure, URCJ-1-e, and the optimized structures, URJC-1-A and –B.

The deformation of the pores is also reflected in properties such as the pore volume and surface area. Table S3 shows the structural properties of the different URJC-1, surface area, pore volume, and framework density. The calculation of accessible pore volume strongly depends on the probe molecule.[81-83] The pore volume obtained experimentally from the argon adsorption isotherm at 87 K is 0.24 cm$^3$/g, while the calculated value of the URJC-1-e using helium is 0.32 cm$^3$/g. To directly compare the values, a correction using the ratio of the Van der Waals radii of the probe molecules is needed, 140 and 188 pm for He and Ar atoms, respectively. Table S3 shows the accessible pore volumes using He and Ar atoms as probe molecules. The most significant difference is found between URJC-1 (as synthetized) and URJC-1-e, showing an increase from 0.24 to 0.26 cm$^3$/g. The relaxed structure shows larger cell lengths and pore volume, but the preservation of the symmetry and the invariability of the surface area make those changes irrelevant in the adsorption behaviour. This finding reinforces the idea that the diamond-shaped pore distortion is the determinant factor. The adsorption of the hydrocarbons produces a change in the pore shape, causing an increase in the URJC-1 (-A and -B) surface area. Figure 9 shows the superposition of the optimized structures URJC-1-A and -B with the reference structure, URJC-1-e. As indicated by the surface areas, the adsorbed paraffins trigger a more pronounced pore distortion. Quantitatively, the variation is shown in the PXRD and pore size distribution (Figure S8).

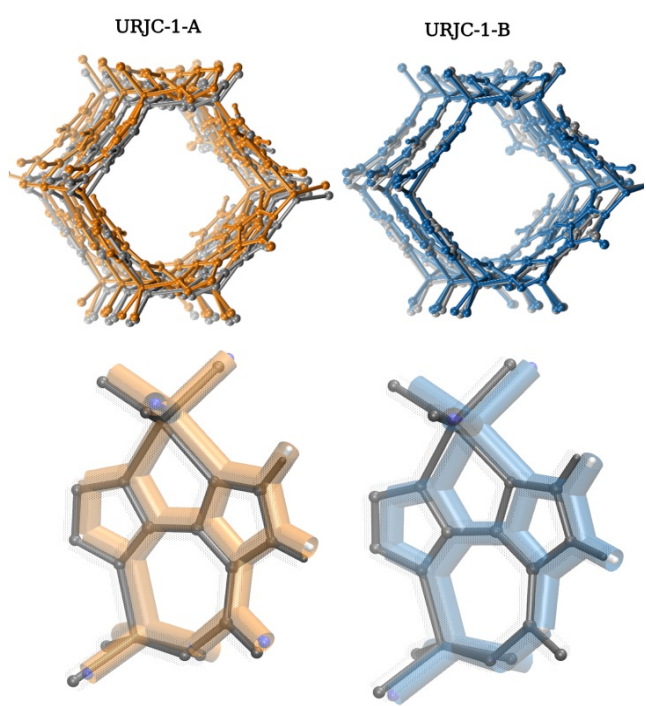

**Figure 9**. Schematic representation of the structure deformation of URJC-1-A (orange) and URJC-1-B (blue) compared with the reference structure, URJC-1-e (gray). Single-pore propagation in c-axis (top) and selected fragment (bottom).

## Conclusions

Two copper-based Metal-Organic Frameworks (Cu-MOF-74 and URJC-1) have been successfully synthesized and characterized. In order to understand their behaviour upon light hydrocarbons adsorption experimental and computational analyses have been performed. Further investigation is needed to determine the potential in gas adsorptive separation processes of these structures.

The results obtained by single-gas adsorption isotherms evidence weaker hydrocarbon-Cu-MOF-74 interactions than the reported in literature for M-MOF-74 analogs. The computational results corroborate the lower impact of Cu in comparison with other metals such as Co, Ni, or Mg, where the difference for alkenes is much more noticeable. In addition, the slightly smaller pore size derives in a lower adsorption capacity.

The results concerning URJC-1 indicate that adsorption occurs in the center of the diffusion channel as a consequence of the narrow pores of the material, providing larger Cu-C distances. Computational analysis reveals the induced framework deformation. A change of the diamond-shaped pores is triggered by the guest molecules affecting the adsorption capacity per adsorbate and the structural properties of the optimized structures. The behavior of URJC-1 is analyzed and reported in this study for the first time.

The adsorption enthalpies reveals a usual behaviour, higher interaction with larger molecules, following the hierarchy C1<C2<C3. Unexpectedly, the effect of the OMS is negligible in the enthalpy of the olefins; thus, no difference is observed between olefins/paraffins pairs with the same chain length in Cu-MOF-74 and only slight differences at low loading are obtained in URJC-1. Based on the adsorption curves and the energy analysis, we can conclude that tuning operation conditions to the optimal pressure range will allow the use of these copper MOFs in light hydrocarbons mixtures separation processes based on chain length. Whereas the capability of these MOFs to separate olefins from paraffins is not clear and needs further investigation, especially in the case of URJC-1, where the induced deformation of the pores is adsorbate-dependent and could affect the behaviour of the performance in the mixture.

## Author Contributions

†ALT (computational) and EAG (experimental measurements) share first authorship with equal contribution.

## Conflicts of interest

There are no conflicts to declare

## Acknowledgments

The authors thank the Spanish Ministry of Science and Innovation and the Spanish State Research Agency for the financial support (Project PGC2018-099296-B-I00). This work is supported by the Irene Curie Fellowship (ICF). E. A.-G thanks MICINN for a Juan de la Cierva Formación fellowship (FJC2019-039015-I) and Margarita Salas fellowship (MS21-035).

# Guest-induced structural deformation in Cu-based Metal-Organic Framework upon hydrocarbon adsorption

Azahara Luna-Triguero,*[†][a,b] Eduardo Andres-Garcia,[†][c,d] Pedro Leo,*[c,e] Willy Rook,[c] Freek Kapteijn[c]

[a] Energy Technology, Department of Mechanical Engineering, Eindhoven University of Technology, P.O. Box 513, 5600 MB Eindhoven, The Netherlands.
[b] Eindhoven Institute for Renewable Energy Systems (EIRES), Eindhoven University of Technology, PO Box 513, Eindhoven 5600 MB, The Netherlands
[c] Catalysis Engineering, ChemE, TUDelft, Van der Maasweg 9, 2629 HZ Delft, The Netherlands.
[d] Instituto de Ciencia Molecular (ICMol), Universidad de Valencia, c/Catedrático José Beltrán, 2, Paterna, 46980, Spain.
[e] Department of Chemical and Environmental Technology, Rey Juan Carlos University, Calle Tulipán s/n, 28933 Móstoles, Spain.

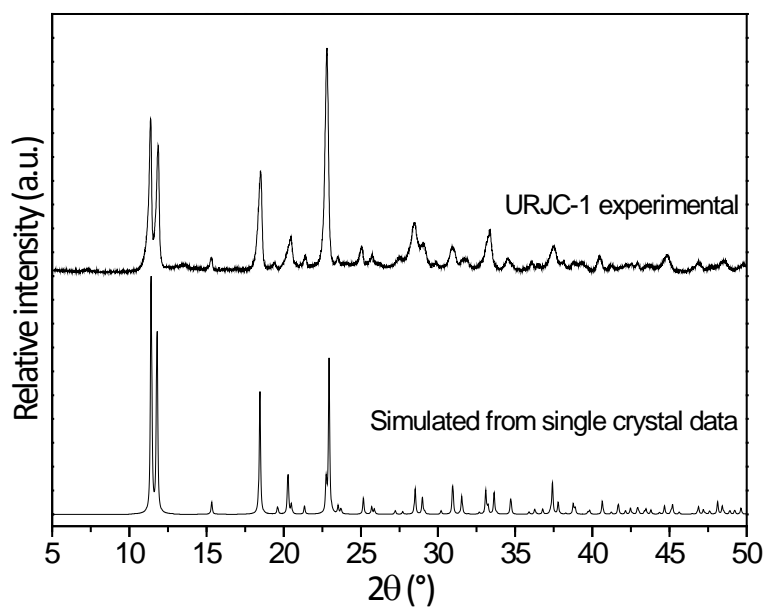

**Figure S1.** XRD patterns of powder URJC-1 sample and simulated from single-crystal data.

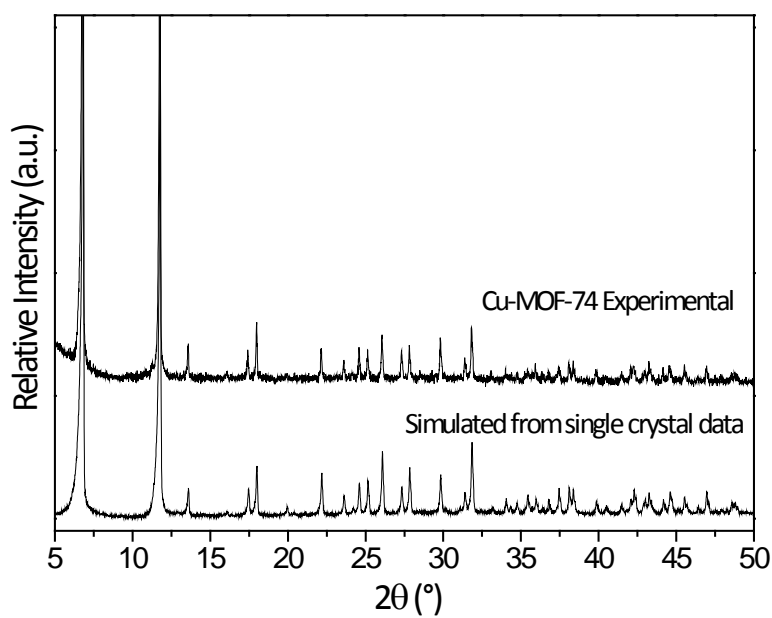

**Figure S2.** XRD patterns of powder Cu-MOF-74 sample and simulated from single-crystal data.

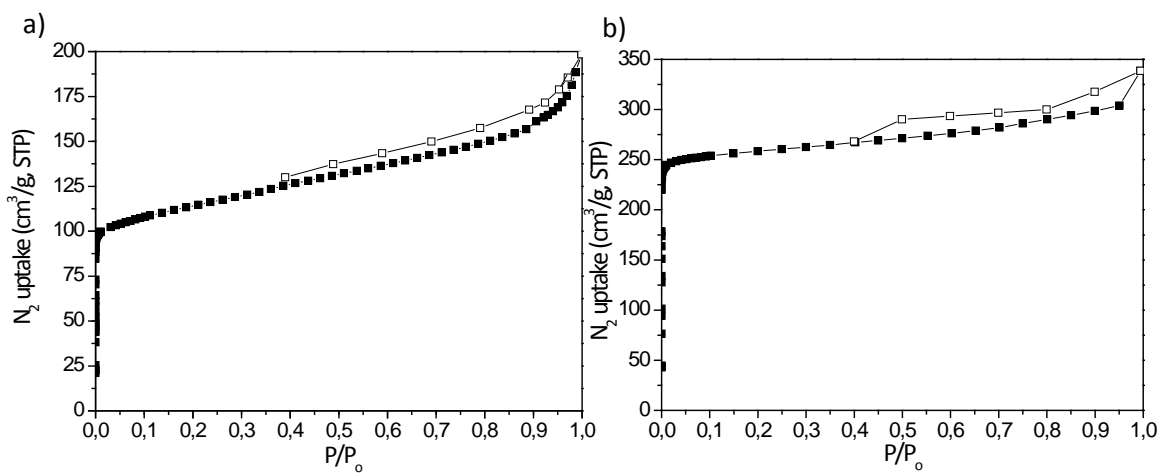

**Figure S3**. Experimental nitrogen adsorption isotherms at 77 K, URJC-1 (a) and Cu-MOF-74 (b).

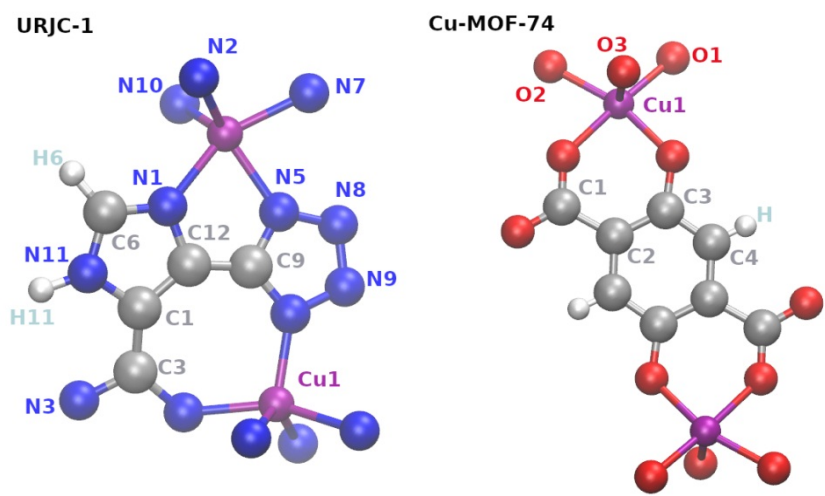

**Figure S4**. Schematic representation of URJC-1 and Cu-MOF-74 linkers and atom labels.

**Table S1**. Point charges.

| Atom Label | URJC-1-e | URJC-1-A | URJC-1-B | Cu-MOF-74 |
|---|---|---|---|---|
| Cu1 | 3.117 | 1.158 | 1.165 | 0.831 |
| N4 | -0.408 | -0.193 | -0.193 | |
| N2 | -0.311 | -0.094 | -0.094 | |
| N3 | -0.345 | -0.093 | -0.093 | |
| N11 | -0.55 | -0.130 | -0.130 | |
| H11 | 0.302 | 0.063 | 0.063 | |
| C12 | -0.061 | 0.016 | 0.016 | |
| C1 | 0.048 | -0.032 | -0.032 | -0.075 |
| C2 | | | | 0.345 |
| C6 | 0.011 | -0.018 | -0.018 | |
| H6 | 0.118 | 0.086 | 0.086 | |
| N1 | -0.512 | -0.106 | -0.106 | |
| C3 | 0.22 | -0.015 | -0.015 | 0.156 |
| C4 | | | | -0.139 |
| N10 | -0.39 | -0.154 | -0.154 | |
| N8 | -0.249 | -0.031 | -0.031 | |
| N7 | -0.384 | -0.069 | -0.069 | |
| N5 | -0.406 | -0.093 | -0.093 | |
| N9 | -0.247 | -0.112 | -0.112 | |
| C9 | 0.047 | -0.182 | -0.182 | |
| O1 | | | | -0.394 |
| O2 | | | | -0.431 |
| O3 | | | | -0.368 |
| H | | | | 0.074 |

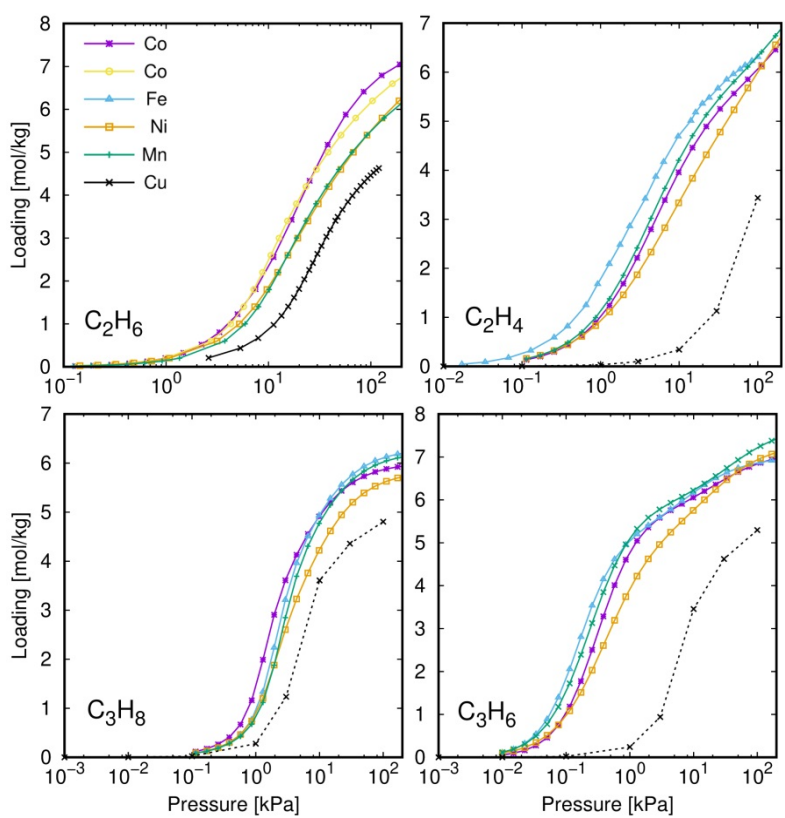

**Figure S5**. Comparison of experimental adsorption isotherms of ethane (298 K), ethene (318 K), propane (318 K), and propene (318 K) in M-MOF-74 (M=Co, Fe, Mn, Ni) taken from literature[1-4] and Cu-MOF-74 measured (solid line) or calculated (dashed line).

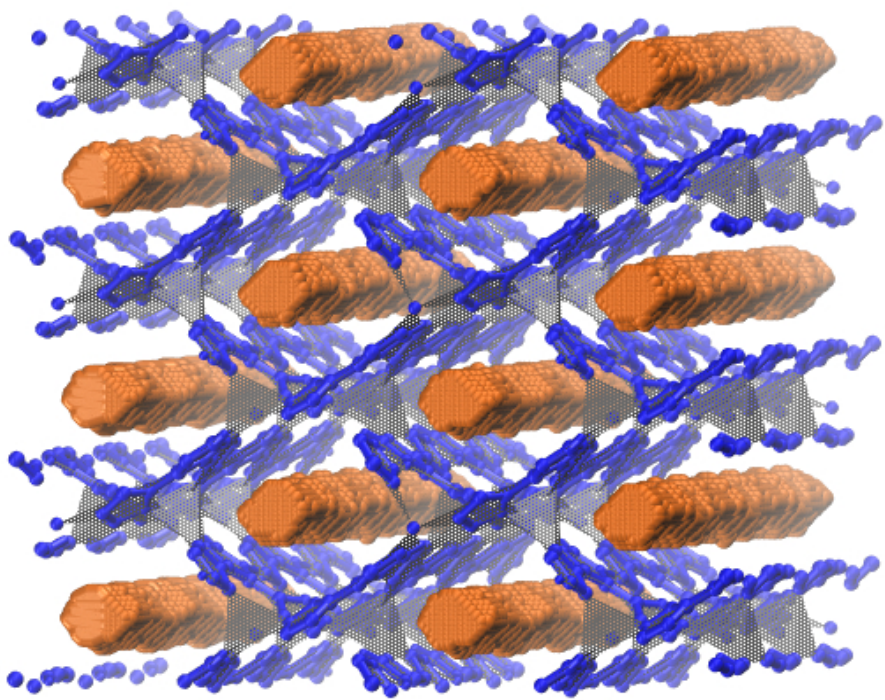

**Figure S6**. Schematic representation of the adsorption sites in URJC-1 using a probe molecule of 3.4 Å.

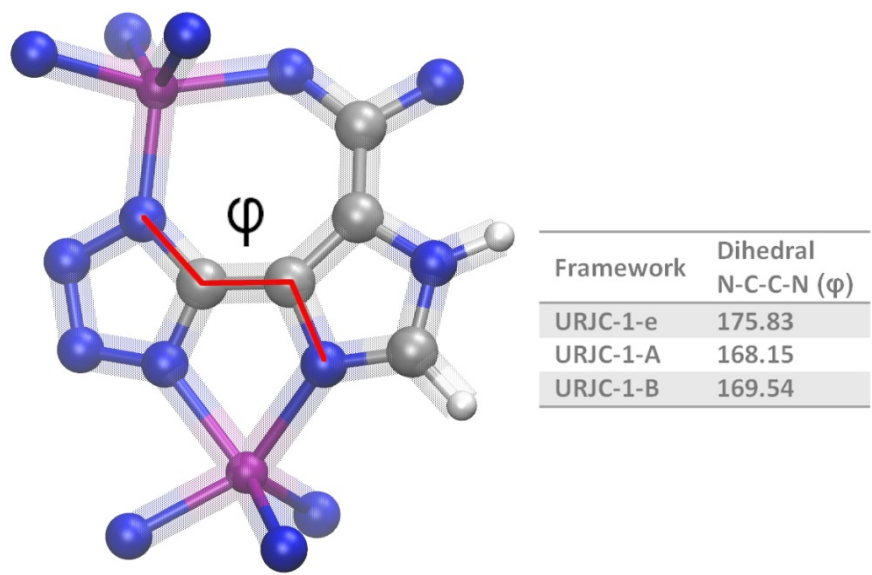

| Framework | Dihedral N-C-C-N ($\varphi$) |
|---|---|
| URJC-1-e | 175.83 |
| URJC-1-A | 168.15 |
| URJC-1-B | 169.54 |

**Figure S7**. Definition and comparison of the dihedral angle (highlighted in red) of URJC-1-e, -A, and –B. Carbon atoms in gray, copper atoms in violet, nitrogen atoms in blue and hydrogen atoms in white.

**Table S2.** Cell parameters of as-synthesized and optimized URJC-1 structures.

| Cell parameter | As-synthesized | URJC-1-e | URJC-1-A | URJC-1-B |
|---|---|---|---|---|
| $a$ [Å] | 25.97 | 27.14 | 26.60 | 26.75 |
| $b$ [Å] | 27.14 | 28.85 | 29.79 | 28.94 |
| $c$ [Å] | 29.99 | 28.29 | 27.68 | 28.15 |
| $\alpha$ | 90 | 90 | 90 | 89.96 |
| $\beta$ | 90 | 90 | 90 | 91.50 |
| $\gamma$ | 90 | 90 | 90 | 90.31 |
| $V_{cell}$ [Å$^3$] | 22148.35 | 21146.14 | 21928.46 | 21787.29 |

**Table S3**. Calculated surface area ($S_A$), available pore volume $V_p$, and framework density of as-synthesized and optimized URJC-1 structures.

| Framework | $S_A$ [m$^2$/g] | $V_p$ [cm$^3$/g] (He) | $V_p$ [cm$^3$/g] (Ar) | $\rho$ [g/cm$^3$] |
|---|---|---|---|---|
| **URJC-1 (as-synthesized)** | 761.48 | 0.32 | 0.24 | 1.50 |
| **URJC-1-e** | 762.28 | 0.35 | 0.26 | 1.43 |
| **URJC-1-A** | 821.64 | 0.32 | 0.24 | 1.43 |
| **URJC-1-B** | 790.73 | 0.31 | 0.23 | 1.46 |

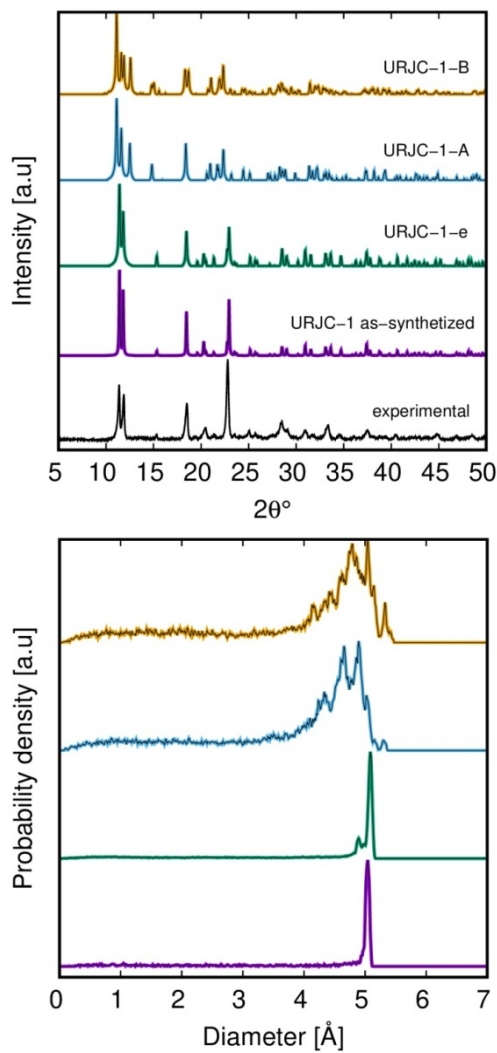

**Figure S8**. Comparison of XRD of as-synthetized URJC-1 (experimental and simulated), and optimized structures from energy minimizations (top) and pore size distribution (bottom) of URJC-1.